# *In situ* investigation of conducting interface formation in LaAlO$_3$/SrTiO$_3$ heterostructure

Hyang Keun Yoo,[1,2,3] Luca Moreschini,[1] Aaron Bostwick,[1] Andrew L. Walter,[1,4] Tae Won Noh,[2,3] Eli Rotenberg[1,*] and Young Jun Chang,[5,6,$]

[1]*Advanced Light Source (ALS), E. O. Lawrence Berkeley National Laboratory, Berkeley, California 94720, USA*

[2]*Center for Correlated Electron Systems, Institute for Basic Science (IBS), Seoul 08826, Korea*

[3]*Department of Physics and Astronomy, Seoul National University, Seoul 08826, Korea*

[4]*Photon Sciences Directorate, Brookhaven National Laboratory, NSLS II, Upton, New York 11973, USA*

[5]*Department of Physics, University of Seoul, Seoul 02504, Korea*

[6]*Department of Smart Cities, University of Seoul, Seoul 02504, Korea*

*erotenberg@lbl.gov & $yjchang@uos.ac.kr

**ABSTRACT**

The high-mobility conducting interface (CI) between LaAlO$_3$ (LAO) and SrTiO$_3$ (STO) has revealed many fascinating phenomena, including exotic magnetism and superconductivity. But, the formation mechanism of the CI has not been conclusively explained. Here, using *in situ* angle-resolved photoemission spectroscopy, we elucidated the mechanisms for the CI formation. In as-grown samples, we observed a built-in potential ($V_{bi}$) proportional to the polar LAO thickness starting from the first unit cell (UC) with CI formation appearing above 3 UCs. However, we found that the $V_{bi}$ is removed by synchrotron ultraviolet (UV)-irradiation; The built-in potential is recovered by oxygen gas ($O_2(g)$)-exposure. Furthermore, after UV-irradiation, the CI appears even below 3UC of LAO. Our results demonstrate not only the $V_{bi}$-driven CI formation in as-grown LAO/STO, but also a new route to control of the interface state by UV lithographic patterning or other surface modification.

Transition-metal-oxide (TMO) heterostructures have been intensively investigated due to the intriguing collective phenomena that occurs at the heterointerface, which is absent in bulk TMOs[1-4]. In particular, the LaAlO$_3$/SrTiO$_3$ (LAO/STO) heterostructure serves as a representative system in terms of its unique interface properties[5-15]. Both bulk LAO and STO are band insulators with gaps of 5.6 eV and 3.2 eV, respectively. However, at their interface, one can observe metallic behavior with high-mobility conducting electrons when LAO approaches a critical thickness $t_c \sim$ 3-4 unit cells (UC)[7,8]. These high-mobility electrons also exhibit exotic two-dimensional superconductivity and coexistent ferromagnetism[9-11]. Additionally, one can write or erase such interesting conducting properties with scanning probes, a desirable quality for next-generation nanoscale electronic devices[12,13].

The mechanism of the conducting interface (CI) formation in LAO/STO is under extensive debate[16-31]. The polarity of LAO has been considered to be an important ingredient in the formation of the CI[16,17]. Particularly, the "polar catastrophe" model suggests that the deposition of LAO films, consisting of alternating charged LaO$^+$ and AlO$_2^-$ layers, onto TiO$_2$-terminated STO induces a divergence of the electric potential. As a result, the conducting charge carriers are redistributed to the interface as the LAO layers become thicker than $t_c$[5,16]. However, experimental results have revealed a negligibly small built-in potential ($V_{bi}$)[25-29], is either not observed[25-27] or found to be much smaller, less than 0.3 eV/UC[28,29], than the value expected in this scenario, ~0.9 eV/UC[30,31]. Other possible mechanisms of the CI have been suggested, such as cation mixing[18,19], oxygen vacancies (O$_V$) at the interface[20,21], and O$_V$ on the LAO surface[22-24]. Nonetheless, the properties of the CI, including $t_c$, cannot be fully understood by these proposed alternatives. Moreover, *ex situ* measurements of the physical properties of polar LAO ultrathin films on STO are one of the obstacles, because the polar surface can change drastically in air[32,33].

Here, by using *in situ* angle-resolved photoemission spectroscopy (ARPES), we elucidated the mechanisms of the conducting LAO/STO heterointerface formation. In as-grown LAO/STO samples, we observed the $V_{bi}$ in LAO accumulates linearly with thickness. Simultaneously, in the STO side, the band bending occurs to compensate the $V_{bi}$ near the interface and the conducting electrons appear when the conduction band minimum (CBM) crosses the Fermi level ($E_F$). The $t_c$ for the CI is 3 UC, consistent with previous reports[5,6]. On the other hand, we found that the $V_{bi}$ can be erased and revived by synchrotron ultraviolet (UV)-irradiation and oxygen gas ($O_2(g)$)-exposure, respectively. Additionally, under UV-irradiation, the CI appears even below 3UC of LAO. This result suggests a new avenue to control of the heterointerface state by UV lithographic patterning or other surface modification.

We prepared the high-quality epitaxial LAO films on defect-free 10-UC-thick STO buffer layer covered single crystal Nb-0.5wt%-doped STO (001) using pulsed laser deposition (PLD)[34-36]. Note that the conducting STO substrate was used to prevent the charging problem during the ARPES experiments. The growth temperature and the laser energy density were 600°C and 1~1.5 J cm$^{-2}$, respectively. The oxygen partial pressure was $1 \times 10^{-4}$ Torr during buffer STO and LAO film depositions, and increased in $1 \times 10^{-2}$ Torr during cooling. The sample consisted of a series of LAO films of discretely increasing thickness (from 1 to 4 UC), grown stepwise on TiO$_2$-terminated STO [Fig. 1(a)]. Each stage of growth (designated L$n$, where $n$=the number of UC grown) was much wider than the ~100 μm wide probe beam, and was grown using a movable shutter synchronized to the observed intensity oscillations of the electron diffraction patterns during growth, as described previously[30]. By growing samples L1-L4 on the same substrate, we could minimize sample-to-sample variations in growth parameters (such as deposition rate, temperature and substrate quality) and therefore ensure that only thickness-related variations in electronic structures are probed.

*In situ* ARPES measurements were conducted in an end-station equipped with ARPES and PLD, sharing the same ultrahigh vacuum envelope, at the Beamline 7.0.1 of the Advanced Light Source. After film deposition, the sample was transferred in vacuo to the analysis chamber with a base pressure of $5 \times 10^{-11}$ Torr. The sample temperature during ARPES was 100~150 K. The total energy resolution (photons + electrons) was around 30 meV; the photon energy was 155 eV in order to probe states near the Γ point of the STO. The measurement power density of the photon source was kept at 0.05 W cm$^{-2}$, with the exception of the UV-irradiation case (3.2 W cm$^{-2}$), to minimize the UV-irradiation effects on the sample surfaces[36,37].

The ARPES spectra of the as-grown samples are shown in Figs. 1(b)-(i-iv) as a function of LAO thickness from 1 to 4 UC (L1 and L4 in the figure, respectively). The L1 and L2 samples show no spectral weight (SW) at the Fermi level ($E_F$). However, the SW at $E_F$ becomes evident for samples of L3 and above. This result indicates that the as-grown LAO/STO sample exhibits conducting behavior above L3, which is consistent with previous reports[7,8]. It is worth noting that cation mixing at the interface were proposed as the possible causes of the conducting behavior in the LAO/STO heterostructure[18,19]. In the present study, this possibility can be ruled out due to the absence of SWs in L1 and L2.

The UV-irradiation experiments were performed on as-grown samples[36-39]. Figure 1(c) shows real-time SW changes at $E_F$ under UV-irradiation. The LAO/STO samples exhibit an increase in the SW at $E_F$, implicating the creation of conducting electrons, even below L3. The ARPES spectra of UV-irradiated samples in Figs. 1(d)-(i-iv) reveal the appearance of the electron band more clearly. This result indicates that the UV-irradiated samples exhibit no $t_c$ regarding conducting behavior, contrary to that of the as-grown samples.

The conducting electrons are expected to be located near the interface between LAO and STO[5-15,40]. However, the possible involvement of surface-localized states for the observed conducting electrons should be ruled out. If the observed electron bands originate from the LAO surface state, there should be no thickness variation of the observed band for either the as-grown or the UV-exposed samples. However, the as-grown samples exhibit an apparent critical LAO layer thickness for conducting behavior. Additionally, the UV-induced SW at $E_F$ is strongly dependent on the LAO thickness; the thicker LAO sample shows a smaller increase in the SW at $E_F$. These experimental results can be explained by the conducting electrons near the interface, not the surface state on LAO. From these results, we concluded that the observed conducting electrons are located near the LAO/STO heterointerface for both the as-grown and the UV-irradiated samples.

To understand the mechanisms of the CI formation, we measured the angle-integrated PES for the valence band maximum (VBM) and the core levels vs. LAO thickness. In the as-grown samples, the Al 2p core level and the VBM show an approximately linear shift with thickness, attributed to the build-up of electrostatic $V_{bi}$ formed inside the LAO layers [Figs. 2(a),(b)][28,29]. A shift of $\Delta V \sim 0.17$ eV/UC was determined, between samples STO-L3. Note that the $V_{bi}$ exhibits saturation behavior with the emerging CI above $t_c$ due to a compensation by conducting electrons. This observed $V_{bi}$ is expected to originate from the stacking of the polar $LaO^+$ and $AlO_2^-$ layers[5,16,28]. Correspondingly, in the STO side, a smaller, opposite shift in the Ti $2p_{3/2}$ core level was observed as $\Delta V \sim 0.05$ eV/UC attributed to band bending in the STO substrate near the interface to compensate the $V_{bi}$ [Fig. 2(c)]. The interfacial conducting electrons appear when the CBM crosses the $E_F$ above $t_c$[41].

On the other hand, effects related to $V_{bi}$ largely disappeared after UV-irradiation. As shown in Fig. 2(d), the Al 2p core-level shift becomes much more or less constant with LAO thickness, indicating the disappearance of $V_{bi}$ in LAO. Additionally, the Al 2p and the VBM [Fig 2(e)] showed a negative shift, indicating a formation of a negative band offset (~0.35 eV) of the LAO compared to that of the STO[25,42]. However, even without $V_{bi}$ in LAO, the Ti $2p_{3/2}$ core-level in Fig. 2(f) exhibits a further shift than that of the as-grown samples, see Supplemental Material, Fig. S1, for direct comparison between the as-grown and the UV-irradiated samples. This implies that band bending is enhanced after UV-irradiation, which results in a stronger SW at $E_F$ as shown in Figs. 1(c),(d).

A possible scenario to explain these experimental results is that UV-irradiation induced oxygen desorption on the LAO surface[37-39]. The $O_V$ on the LAO surface provides extra electrons, which induce compensation of $V_{bi}$ in the LAO layers[22-24]. Additionally, the surface electrons from the $O_V$ can drive the observed band bending [37-39]. Note that, in Fig. 1(d), we did not observe an additional defect state associated with the $O_V$ on the LAO surface. This suggests that the $O_V$ defect state formed above $E_F$, and the extra electrons from the $O_V$ are transferred to the band-bent Ti states at $E_F$[22-24]. To verify this hypothesis, especially, the creation of $O_V$ on the LAO surface by UV-irradiation, we investigated the change in core-level spectra during UV- and $O_2(g)$-exposures[39]. Figure 3 shows real-time O 1s and Al 2p core-level spectra for the L4 sample. Both peak shifts indicate that $O_2(g)$-exposure fully or nearly fully compensates the UV-irradiation effects. This experimental result strongly supports our proposal.

Band diagrams of the conducting LAO/STO heterointerfaces, determined by our PES experiments, are illustrated in Fig. 4. Without UV-irradiation, $V_{bi}$ in LAO appears (~0.17 eV/UC) and the $E_F$ of STO is located at around 3.1 eV. Band bending occurs in the STO side to compensate the $V_{bi}$ and, above L3, the CBM crosses $E_F$ and the interface becomes conducting. On the other hand, with UV-irradiation, the $V_{bi}$ within the LAO layer disappeared due to the compensating potential induced by surface $O_V$. Furthermore, the extra electrons from the $O_V$ accumulate in the CBM at $E_F$, because the defect state on the LAO surface is located above $E_F$. As a result, after UV-irradiation, the CI is present, even below L3. The UV-induced oxygen desorption on the LAO surface can be recovered by $O_2(g)$-exposure, which revives $V_{bi}$ within the LAO layers.

The emerging conducting electrons at the LAO/STO heterointerface, possibly related to the precise roles of both nonpolar STO and polar LAO, cation mixing, $O_V$ at the interface, and $O_V$ on LAO surface, has been remained an open question[16-31]. Under controlled *in situ* condition, we determined that the $V_{bi}$ indeed drives the formation of the CI state in the as-grown LAO/STO heterostructure. In addition, it is possible to introduce a second factor, through photon-irradiation, which reduces the $t_c$ for CI formation down to 1 UC with $O_V$ on the LAO surface. These results will shed light on the understanding of the CI at LAO/STO heterostructure.

We should mention the possible microscopic origin of the conducting electrons in the as-grown LAO/STO samples. Although we speculated that UV-induced $O_V$ favors the CI state formation, it could just as well be that the as-grown samples also have $O_V$[5,6,20,21]. There is no SW at $E_F$ in buffer STO, but the in-gap state is observed which might originate from $O_V$ [Supplemental Material, Fig. S3][37]. The $E_F$ of STO at around 3.1 eV also implies the possible $O_V$ existence[41]. Additionally, the $O_V$ distribution in STO can be changed to compensate the $V_{bi}$ in LAO in high-temperature growth condition[43-45]. Namely, the $O_V$ concentration becomes higher near the interface to reduce the internal electric field, which could result in the band bending in STO side [Fig. 2(c)] and the reduction of the $V_{bi}$ to 0.17 instead of 0.9 eV/UC. Probing the oxygen core levels could clarify the existence and the redistribution of the $O_V$ in principle [Supplemental Material, Fig. S2], but it is still complicated by the presence of O atoms in both the film and the substrate, of which near-surface coherent Bragg rod analysis is further requested.[46]

In conclusion, we performed *in situ* ARPES measurements on the LAO/STO heterostructure to elucidate the underlying mechanism for the CI formation. We demonstrated that the CI can be created by polarity-induced $V_{bi}$ in the LAO layer of the as-grown samples or by UV-irradiation-generated $O_V$ on the LAO surface. Our results show the advantage of combining state-of-the-art film fabrication techniques with *in situ* ARPES measurements for unveiling the intriguing properties of the TMO heterointerface. Furthermore, a clear implication of our work is that the interface electronic properties can be radically tuned by control of the surface of heterostructures.

**ACKNOWLEDGEMENTS**

This work was supported by the Basic Study and Interdisciplinary R&D Foundation Fund of the University of Seoul (2019) for Y.J. Chang. This work was supported by the National Research Foundation (NRF) grants funded by the Korean government (No. NRF2019K1A3A7A09033389). The Advanced Light Source is supported by the Director, Office of Science, Office of Basic Energy Sciences, of the U.S. Department of Energy under Contract No. DE-AC02-05CH11231. This work was supported by IBS-R009-D1. Part of this study has been performed using facilities at IBS Center for Correlated Electron Systems, Seoul National University. L.M. acknowledges support by a grant from the Swiss National Science

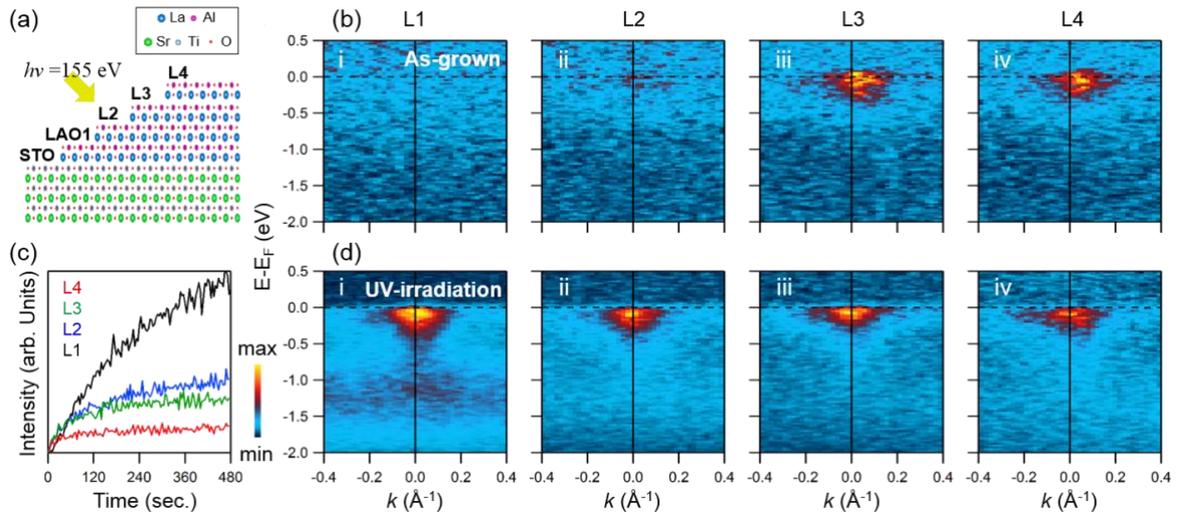

Fig. 1. (a) Schematic diagram for experimental sample geometry. (b)-(i~iv) ARPES spectra of as-grown samples with varying the LaAlO$_3$ (LAO) thickness from 1 (L1) to 4 (L4) unit cells (UC). The spectral weight (SW) at Fermi level ($E_F$) appears above 3 UC of LAO, which indicates the formation of the conducting interface. (c) The real-time intensity change in the SW at $E_F$ with ultraviolet (UV)-irradiation. (d)-(i~iv) ARPES spectra of UV-irradiated samples. All of the samples show the SW at $E_F$.

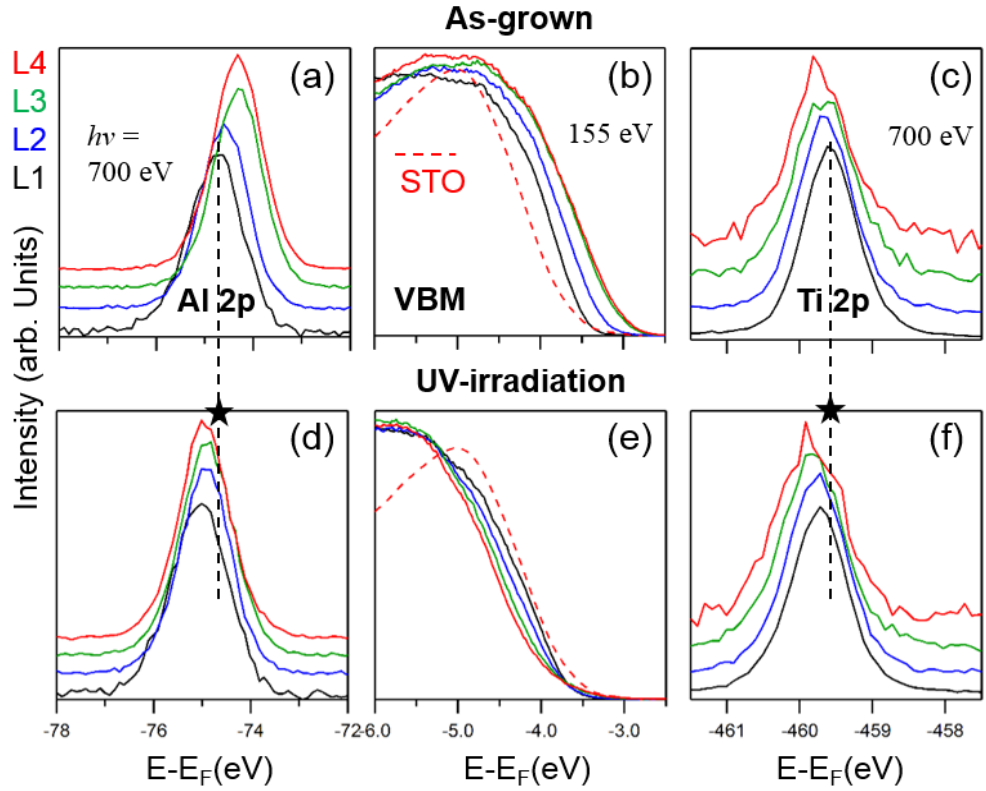

Fig. 2. Integrated photoemission spectra of Al 2p, valence band maximum (VBM), Ti $2p_{3/2}$ for (a)-(c) as-grown and (d)-(f) UV-irradiated samples. In the as-grown samples, the Al 2p in (a) and the VBM in (b) have peak shifts toward lower binding energy proportional to the LAO thickness, i.e. ~0.17 eV/UC. Simultaneously, the Ti $2p_{3/2}$ shows a peak shift, i.e. 0.05eV/UC, to higher binding energy in (c). On the other hand, in the UV-irradiated samples, the Al 2p in (d) becomes much more or less constant with LAO thickness. Additionally, the Al 2p in (d) and the VBM in (e) showed a negative shift compared to that of the as-grown samples, which indicates the appearance of the $V_{offset}$ ~0.35 eV. However, the Ti $2p_{3/2}$ in (f) exhibits a further shift than that of the as-grown samples. For comparison with as-grown L1, its peak position is marked by a black asterid in (d) and (f).

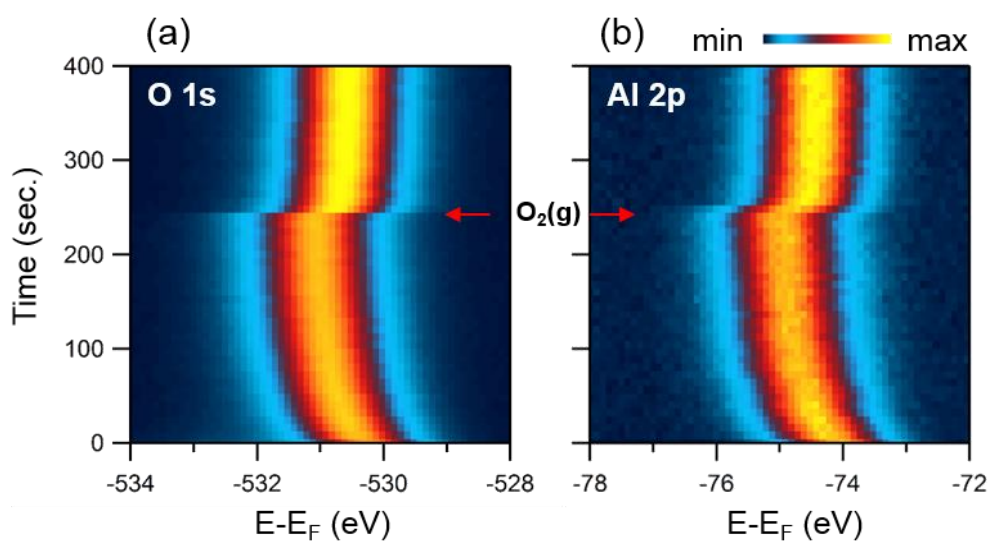

Fig. 3. Real-time (a) O 1s and (b) Al 2p core-level spectroscopy of the L4 sample. UV-irradiation effects are observed, as evidenced by the peak shift to a higher binding energy. The sample was then exposed to $O_2(g)$, which is indicated in the figure by red-arrows. The oxygen partial (base) pressure was $1 \times 10^{-8}$ ($5 \times 10^{-11}$) Torr. The peak shift is the opposite direction, which indicates compensation of the UV-irradiation effects by $O_2(g)$-exposure.

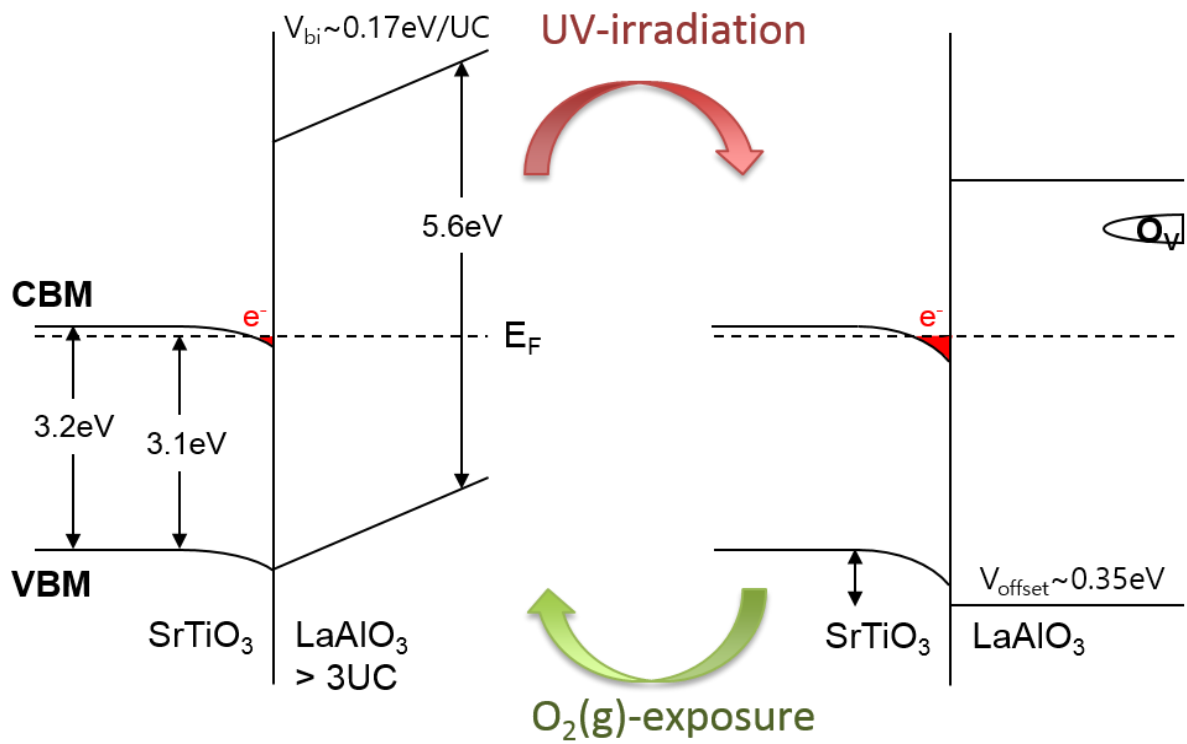

Fig. 4. Band diagrams of the metallic LAO/SrTiO$_3$ interface determined by the present experiments. The as-grown and UV-irradiated samples are depicted in left and right, respectively. The details are described in the main text.